# Efficient indoor *p-i-n* hybrid perovskite solar cells using low temperature solution processed NiO as hole extraction layers

Lethy Krishnan Jagadamma, Oskar Blaszczyk, Muhammad T. Sajjad, Arvydas Ruseckas, Ifor D. W. Samuel*

Organic Semiconductor Centre, SUPA, School of Physics & Astronomy, North Haugh, St Andrews, KY16 9SS, United Kingdom

E-mail: idws@st-andrews.ac.uk

**Abstract:**

Hybrid perovskites have received tremendous attention due to their exceptional photovoltaic and optoelectronic properties. Among the two widely used perovskite solar cell device architectures of *n-i-p* and *p-i-n*, the latter is interesting in terms of its simplicity of fabrication and lower energy input. However this structure mostly uses PEDOT:PSS as a hole transporting layer which can accelerate the perovskite solar cell degradation. Hence the development of stable, inorganic hole extraction layers (HEL), without compromising the simplicity of device fabrication is crucial in this fast-growing photovoltaic field. Here we demonstrate a low temperature (~100 °C) solution - processed and ultrathin (~ 6 nm) NiO nanoparticle thin films as an efficient HEL for $CH_3NH_3PbI_3$ based perovskite solar cells. We measure a power conversion efficiency (PCE) of 13.3 % on rigid glass substrates and 8.5 % on flexible substrates. A comparison with PEDOT:PSS based $MAPbI_3$ solar cells (PCE ~ 7.9 %) shows that NiO based solar cells have higher short circuit current density and improved open circuit voltage (1.03V). Apart from the photovoltaic performance under 1 Sun, the efficient hole extraction property of NiO is demonstrated for indoor lighting as well with a PCE of 23.0 % for NiO based $CH_3NH_3PbI_{2.9}Cl_{0.1}$ *p-i-n* solar cells under compact fluorescent lighting. Compared to the perovskite solar cells fabricated on PEDOT:PSS HEL, better shelf-life stability is observed for perovskite solar cells fabricated on NiO HEL. Detailed microstructural and photophysical investigations imply uniform morphology, lower recombination losses, and improved charge transfer properties for $CH_3NH_3PbI_3$ grown on NiO HEL.

1. Introduction

Emerging hybrid perovskite solar cells hold the promise for inexpensive and highly efficient photovoltaic technology, with certified power conversion efficiencies of 24.2 %[1]. The most successful and widely investigated perovskite solar cell architecture is *n-i-p* consisting of high temperature processed planar and mesoporous $TiO_2$ as electron transporting layer (ETL) and expensive Spiro OMeTAD as hole transporting layer[2, 3]. In addition to the high energy input, the *n-i-p* architecture suffers serious J-V hysteresis and Spiro OMeTAD is a poor barrier for the migration of Au/Ag top metal contacts, critically degrading the performance of perovskite solar cells[4]. The alternate and emerging *p-i-n* planar architecture is promising in terms of simplifying fabrication, reducing energy input in terms of interlayer and reducing the detrimental hysteresis effect[5]. Because of these attributes, this device architecture is favourable for roll-to roll fabrication of perovskite solar cells on flexible and non-customized substrates. However, this architecture has challenges, such as lower efficiency and faster degradation due to the most commonly employed hole transporting layer of PEDOT:PSS, which is hygroscopic and acidic in nature[6]. Development of solution processed inorganic and stable hole extraction layers for the planar *p-i-n* structure is therefore desirable to enable simple, energy efficient and cost effective fabrication of perovskite solar cells.

A crucial aspect of the operation of perovskite solar cells is the presence of charge selective contacts. For their efficient functioning, these layers need to meet important attributes of high transparency in the visible range, energy level matching with the perovskite active layer, good charge transport properties and being easy to deposit. Among the p-type inorganic semiconductors, NiO is interesting for hole transporting layers because of its wide bandgap, stability, high hole mobility in bulk, good electron blocking ability and energy level matching with hybrid perovskites such as $CH_3NH_3PbI_3$ [7, 8]. NiO as hole extraction layer has been previously reported for *p-i-n* planar architecture [6, 9-17] and very recently for *n-i-p* perovskite solar cells [18-21]. Power conversion efficiency as high as ~ 20 % has been demonstrated for *p-i-n* perovskite solar cells using NiO as HEL and ~ 11% for *n-i-p* perovskite solar cells. In most of these reports, the NiO is prepared by physical vapor deposition technique such as ALD or sputtering, or by solution processing methods which involve high temperature (270- 500 ºC) annealing. However to exploit the promises for device flexibility and conformability with perovskite solar cells, the interlayers also needs to be fabricated at moderate temperatures and specifically, below 150°C in the case where polyethylene terephthalate (PET) is used as the flexible substrate [22]. So far there are only a few studies on low temperature solution processed NiO HEL for perovskite solar cells. Liu *et al* [21] reported 15.9 % PCE for two step synthesised CH$_3$NH$_3$PbI$_3$ *p-i-n* perovskite solar cells using NiO thin films annealed at 150 ºC and followed by a UV-ozone treatment. Very recently He *et al* [23] and Chen *et al*[24] reported 15-20 % PCE for CH$_3$NH$_3$PbI$_3$ *p-i-n* perovskite solar cells using room temperature processed NiO and Cu:NiO as HEL. Although interesting, these reports do not explore the applicability of using NiO HEL to perovskite compositions other than CH$_3$NH$_3$PbI$_3$, or the shelf-life stability of the fabricated *p-i-n* solar cells. Furthermore, they do not directly compare the perovskite solar cell device performance with the widely used hole extraction layer of PEDOT:PSS.

Hybrid perovskite solar cells have so far been mainly investigated for their performance under outdoor solar illumination with very few reports discussing the *indoor* photovoltaic performance of perovskite solar cells [25-29]. The tuneable bandgap (1.2-3.1 eV) of hybrid perovskites[30] and visible spectral range of indoor light sources give hybrid perovskites considerable potential for indoor light harvesting. Indoor lighting in residential buildings and offices are dominated by compact fluorescent lamps (CFL) and white light emitting diodes (LEDs) which are entirely different in intensity and spectral content compared to the solar spectrum. Indoor solar cells are very interesting for powering smart autonomous systems such as Internet of Things (IoT) applications. IoT is a huge technology field of 'smart' connected devices and is expected to grow rapidly over the coming years. By 2025, IoT is expected to connect 30 billion objects, sensors being more than 60% of devices [31]. Considering the huge number of objects being connected in IoT, the use of power distribution wires or battery is complex, expensive and an environmental issue. Interestingly these wireless sensors need only microwatt ranges of power which could potentially come from indoor solar cells. Hence the development of cost effective, photovoltaic technology to power these sensors is important and hybrid perovskites are an ideal choice.

In the present work, a low temperature (100 $^0$C), solution processed, NiO nanoparticle based thin film as efficient hole extraction layer is demonstrated for CH$_3$NH$_3$PbI$_3$ and mixed halide perovskites such as CH$_3$NH$_3$PbI$_{2.9}$Cl$_{0.1}$ and CH$_3$NH$_3$PbI$_{2.9}$Br$_{0.1}$. These NiO nanoparticles were synthesised using flame spray synthesis method which is capable of producing uniformly distributed metal oxide nanoparticles at a high production rate (200 g/h) in a cost efficient manner and without forming any harmful by-products [32]. Thus these NiO nanoparticles are attractive towards the industrial fabrication of perovskite solar cells. Detailed investigation of the microstructural, optoelectronic and photophysical properties of the CH$_3$NH$_3$PbI$_3$ on NiO /PEDOT:PSS HELs is carried out and compared. Improved microstructural and photophysical

 3

properties of $CH_3NH_3PbI_3$ films grown on NiO hole transporting layers in comparison to those on PEDOT:PSS results in higher (13.3 %) photovoltaic power conversion efficiency for the NiO based perovskite solar cells in comparison to PEDOT:PSS (7.9 %) as HEL. In addition to the performance of these hybrid perovskite solar cells under AM 1.5 G solar spectrum, their photovoltaic performance is characterized for indoor lighting applications. Based on this, a PCE of 23.0 % is demonstrated for NiO based $CH_3NH_3PbI_{2.9}Cl_{0.1}$ *p-i-n* solar cells. Moreover, compared to the perovskite solar cells fabricated on PEDOT:PSS HEL, better shelf-life stability is observed for perovskite solar cells fabricated on NiO HEL. Detailed microstructural and photophysical investigations imply uniform morphology, lower recombination losses, and improved charge transfer properties for $CH_3NH_3PbI_3$ grown on NiO HEL compared to that on PEDOT:PSS.

## 2. Experimental details

*2.1 Preparation of NiO thin films*

The dry NiO nanopowder was prepared by flame spray synthesis using a precursor solution of nickel salt in 2-ethylhexanoic acid; the solution was diluted with 30 wt% THF. The precursor then was fed to a spray nozzle, dispersed by oxygen and ignited by a premixed methane-oxygen flame ($CH_4$: 1.2 l min$^{-1}$ and $O_2$: 2.2 l min$^{-1}$). The off-gas was filtered through a steel mesh filter (20 micron mesh size) by a vacuum pump at about 20 m$^3$h$^{-1}$. The obtained dry nanopowder was collected from the filter mesh. A stable suspension was prepared by dispersing 2.5 wt% nanopowder in ethanol by using an undisclosed dispersant (proprietary information of Avantama AG).

NiO nanoparticle thin films were prepared from the above mentioned NiO nanoparticle suspension (Avantama, P-21, product number 10128, 7 nm particle size) from Avantama. The as-received NiO nanoparticle was ultrasonicated for 1 minute followed by diluting (1:10 by volume) the suspension in anhydrous ethanol. The diluted NiO nanoparticle suspension was sonicated for 5-10 minutes before spin coating. The spin coating was carried out under ambient conditions, at 2000 rpm for 60 seconds. After spin coating, the glass/ITO/NiO substrates were annealed at 100 ºC for 10 minutes.

*2.2 Fabrication of perovskite solar cells*

The pre-patterned ITO-coated glass substrates were cleaned in detergent (sodium dodecyl sulphate-SDS), successively ultrasonicated in deionized water, acetone and isopropyl alcohol, and exposed to oxygen Plasma Asher for 3 minutes. The solution to make the perovskite active layer was prepared by mixing 461 mg of $PbI_2$ (Alfa Aesar 99.999%) and 159 mg of $CH_3NH_3I_3$ (Dyesol) in a mixed solvent of Dimethyl sulfoxide (66 µL) (DMSO, Sigma Aldrich, anhydrous, 99.9%) and Dimethyl formamide (636 µL) (DMF, Sigma Aldrich, anhydrous, 99.9%). The precursor solution was stirred for 1 hour at room temperature and then spin-coated onto the glass/ITO/NiO substrates at 4000 rpm for 30 seconds. During the first 7 s of the spin-coating process, a diethyl ether wash (700 µL) was carried out. After completing the spin coating process, the perovskite film was annealed in vacuum (~ 100 mbar) at 100 ºC for 1 minute followed by annealing in $N_2$ ambient in the glovebox for another 2 minutes. The electron transporting layer of $PC_{61}BM$ was prepared by mixing 10 mg/mL in chlorobenzene and followed by a spin coating process of 1000 rpm for 60 s. The glass/ITO/NiO/perovskite/$PC_{61}BM$ structure was then thermally annealed at 80 ºC for 5 minutes. The perovskite solar cell was then completed by thermally evaporating 1 nm of LiF and 100 nm of Ag contact. In the case of PEDOT:PSS HEL based solar cells, the HEL was prepared by spin coating the CLEVIOS™ P VP AI 4083 P (Heraeus) PEDOT:PSS at 4000 rpm for 60 s and annealing in air at 140 ºC for 10 minutes. All the weighing of the perovskite materials, precursor solution preparation, stirring, perovskite active layer spin coating and the $PC_{61}BM$ layer deposition were performed inside a nitrogen filled glove box.

After the electrode deposition, the devices were encapsulated with a UV optical adhesive and a glass coverslip. The current-voltage characteristics were determined under an illumination intensity of 100 mW/cm$^2$ in air using an air mass 1.5 global (AM 1.5G) Sciencetech solar simulator (SS150-AAA) and a Keithley 2400 source-measure unit. The illumination intensity was verified with a calibrated monosilicon detector and a KG-5 filter. The solar cells were applied a bias sweep from -0.20 V to 1.5 V, with a voltage step of 10 mV and a delay time of 200 ms. The external quantum efficiency (EQE) measurements were performed at zero bias by illuminating the device with monochromatic light supplied from a Xenon arc lamp in combination with a dual-grating monochromator. The number of photons incident on the sample was calculated for each wavelength by using a silicon photodiode calibrated by the National Physical Laboratory (NPL). For the indoor measurements, the LED light source used was Cree (Cree XML T6) and the illuminance intensity (2.2 mW/cm$^2$) was adjusted by changing the input voltage. A fluorescent lamp (RS components, PL 11 lamp, 11 W) was also used as an illumination source (intensity 0.32 mW/cm$^2$). The irradiance level is measured using RK 5710 power radiometer and Optometer P9710.

*2.2 Characterization of the NiO and hybrid perovskite thin films*

The UV-Vis absorption and transmission spectra of NiO and hybrid perovskite thin films were recorded using a Cary 300 spectrometer for the wavelength range of 300-800 nm. The thickness of the films were measured using Dektak II profilometer (stylus force 6 mg and a measurement range of 6.5 µm). HOMO levels of the NiO and PEDOT:PSS films were measured using ambient photoemission (KP Technology APS03 instrument). The Kelvin probe tip has a gold-alloy coating (2 mm diameter) and when used in contact potential difference (CPD) mode, vibrates at 70 Hz at an average height of 1 mm from the sample surface. For recording the ambient photoemission spectrum (APS), the sample was illuminated with a 4-5 mm diameter light spot from a tunable monochromated $D_2$ lamp (4–7 eV). The raw photoemission data were corrected for detector offset; intensity normalized, then processed by a cube root power law. The energy resolution in APS mode is 50-100 meV. The Kelvin probe contact potential difference (CPD) technique was used to measure the surface photovoltage (SPV). SPV is estimated using the relation of [CPD(dark) – CPD(light)]. For the SPV measurement the sample was illuminated using a QTH lamp. Steady state PL spectra were measured with a Hamamatsu streak camera C10910-05, with S-20ER photocathode and 650 nm as excitation wavelength from an optical parametric amplifier, pumped by Pharos laser (from Light Conversion). Time-resolved photoluminescence was spectrally integrated in the range of 720-800 nm and measured with a 3 ps time resolution, using a Hamamatsu streak camera in synchroscan mode, following excitation by 200 fs pulses at 640 nm from an optical parametric amplifier, pumped by Pharos regenerative amplifier at 100 kHz repetition rate. The incident excitation fluence was kept low at 340 nJ cm$^{-2}$, to avoid non-geminate charge recombination. The surface morphology of the NiO and PEDOT:PSS thin films were characterized using atomic force microscopy (AFM). AFM images were obtained with a Bruker MultiMode 8 instrument in the tapping mode. NANOSENSORS™ PPP-NCSTR Si cantilever tips with force constant of 6–7 Nm$^{-1}$ were used as AFM probes. The transmission electron microscopy (TEM) and high resolution TEM were taken using *Titan Themis 200* scanning transmission electron microscope (S/TEM), equipped with an X-FEG Schottky field emission gun and Super-X high sensitivity windowless EDX detector for rapid compositional analysis. The sanning electron microscopy (SEM) measurements of the hybrid perovskite layers were taken using a Hitachi S4800 scanning electron microscope. X-ray diffraction spectra of the $CH_3NH_3PbI_3$ thin films were collected on a Rigaku Miniflex 600 diffractometer using CuK$\alpha_{1,2}$ radiation and a D/tex Ultra detector. Data were collected in the range 5º -70º 2θ with a step size of 0.02º and a time per step of 0.28 s.

## 3. Results and Discussion:

### *3.1 Properties of NiO films and its comparison with PEDOT:PSS*

The characterization of the prepared NiO thin films were done to check the potential of these films to function as hole transporting layers in inverted perovskite solar cells. The transmission properties, the surface morphology, energy level alignment and crystalline properties of the prepared NiO thin films were investigated in this regard. The transmission spectrum of the prepared NiO thin films on ITO substrate is shown in Figure 1(a). A comparison of the PEDOT:PSS transmission spectrum in the visible range is also given. The prepared thin film of NiO is highly transparent and does not further reduce the transmission properties of the ITO substrate in the visible range. The average transmission of the glass/ ITO/NiO thin films are above 80 % and this can be attributed to the high optical band gap (3.6 eV) for NiO thin films. The surface morphology of the NiO nanoparticle thin films on ITO surface obtained by atomic force microscopy is shown in Figure 1(b). Very fine nanoparticles with uniform size distribution and dense packing with full surface coverage is observed. The RMS surface roughness of the corresponding NiO thin film is ~ 3.8 nm. The thickness of the NiO thin film is determined by Spectroscopic ellipsometry method and is ~ 6 nm. To ensure that the NiO film is covering the ITO substrate, AFM imaging of the ITO substrate is also done and is shown in Figure S1(a). The ITO substrate is smoother than NiO nanoparticle film and the RMS roughness is around 1.5 nm. Figure S1(b) shows the surface morphology of the PEDOT:PSS films and is even smoother (RMS roughness of ~ 1.1 nm) than the ITO surface. To study the particle size distribution and crystalline properties, high resolution transmission electron images of the NiO nanoparticle thin film were obtained by scratching particles from the ITO substrate and is shown in Figure 1(c). These NiO nanoparticles have particle size of ~ 5 nm and show clear lattice fringes, indicating their crystalline quality. The histogram of the particle size distribution is shown in Figure S1(c). The selective area diffraction pattern (SAED) given in the inset of figure 1(c) shows that the NiO nanoparticles are highly polycrystalline. The elemental composition of the NiO nanoparticles shown in the TEM image is confirmed by taking the energy dispersive spectra (EDS) from the same imaging area and is shown in Figure S1(d). Only Ni and O elements were observed from the EDS spectrum except Cu coming from the grid. For the efficient charge transport from the perovskite active layer to the charge transporting layer, ohmic contact is needed at the perovskite active layer/electrode interface. The work function and valence band (VB) position of the prepared NiO films were studied using Kelvin probe (KP) and Ambient photoemission spectroscopy (APS). The APS spectra of NiO thin films on ITO substrate is shown in Figure 1(d) and the estimated VB is at 5.32 eV and the measured Fermi level is around 5.26 eV. In order to understand how much the VB of NiO differs from the HOMO level of the most commonly employed hole extraction layer of PEDOT:PSS, the Fermi level and HOMO level of PEDOT:PSS was also measured as shown

 7

in Figure S2. The HOMO level of PEDOT:PSS is 5.46 eV and the Fermi level is 5.29 eV. Compared to NiO nanoparticle films, the PEDOT:PSS HOMO is slightly deeper.

*3.2 Comparison of Photovoltaic properties*

After characterizing the various optoelectronic and structural properties, the low temperature solution processed NiO nanoparticle thin films were tested as hole extraction layers in inverted perovskite solar cells using $CH_3NH_3PbI_3$ as the active layer. The inverted device configuration used consists of glass/ITO/NiO/$CH_3NH_3PbI_3$/$PC_{61}BM$/LiF/Ag and is shown in Figure 2(a). The influence of NiO nanoparticle film thickness on photovoltaic properties of inverted perovskite solar cells is given in Figure S3. The NiO films with thickness around 6 nm show the best photovoltaic properties. The control devices were fabricated using the same device configuration using PEDOT:PSS as the HEL. A comparison of the light J-V characteristics under AM1.5 G solar simulator spectrum for inverted perovskite solar cells using PEDOT:PSS and NiO nanoparticle thin films are shown in Figure 2(b). The corresponding photovoltaic performance parameters are listed in **Table 1.** Compared to the inverted solar cells using PEDOT:PSS as HEL, which show a maximum power conversion efficiency of 7.9 %, the inverted perovskite solar cells with NiO as HEL show much improved photovoltaic performance with a maximum power conversion efficiency of 13.3 %. The consistency and reproducibility of the NiO nanoparticle thin films as an efficient hole extraction layer has been confirmed by fabricating many batches of inverted perovskite solar cells. The histogram distribution of power conversion efficiency is shown in Figure 2(c) and is compared to that of PEDOT:PSS HEL based solar cells. If we compare the photovoltaic properties listed in Table 1, it is obvious that, $CH_3NH_3PbI_3$ perovskite solar cells on NiO HEL have higher open circuit voltage and improved short circuit current density. A comparison of the EQE spectra of the $CH_3NH_3PbI_3$ solar cells based on both NiO and PEDOT:PSS is given in Figure 2(d) and confirms the difference in short circuit current density from the two device configurations using PEDOT:PSS and NiO as HEL.

Since the processing temperature of the NiO hole extraction layer is compatible with flexible substrates such as PET, $CH_3NH_3PbI_3$ perovskite solar cells were fabricated on PET/ITO/NiO substrates. The maximum power conversion efficiency obtained is 8.5 %. The J-V characteristics, EQE spectra and the corresponding photovoltaic performance parameters are listed in **Figure 3(a), (b) and Table 2**. Compared to the $CH_3NH_3PbI_3$ perovskite solar cells fabricated on Glass/ITO/NiO, the solar cells on flexible substrates have slightly lower PCE.

 

The lower PCE is mainly due to the lower FF for the PET/ITO/NiO based perovskite solar cells compared to that on the glass counterpart. This could be related to the higher sheet resistance (60 ohm/sq.) of the PET/ITO substrate compared to glass/ITO substrate (15 ohm/sq).

So far the fabrication of $CH_3NH_3PbI_3$ perovskite solar cells on different HEL layers and the glass/flexible substrate was discussed. Now, in order to understand the differences in solar cell device performance for the two hole extraction layers (NiO and PEDOT:PSS), J-V characteristics of the solar cells were measured as a function of light intensity. By studying the variation of $J_{sc}$ and $V_{oc}$ as a function of light intensity, different charge extraction/recombination mechanism of photogenerated charges prevalent in these devices can be understood. The short circuit current density $J_{sc}$ depends upon the light intensity $I$ by a power law relation of,

$$J_{sc} \propto I^{\alpha}$$

where $\alpha$ is a recombination factor and is unity when all the separated charges are collected by the electrodes. A small decrease of $\alpha$ from unity corresponds to weak bi-molecular recombination losses. Values of $\alpha$ in the range of 0.7-0.8, however, corresponds to space charge effects [33, 34]. The dependence of short circuit current density $J_{sc}$ on light intensity for perovskite solar cells using PEDOT:PSS and NiO as HEL is shown in Figure 4 (a). The nearly straight line in a log-log plot indicates a power law dependence ($J_{sc} \propto I^{\alpha}$). For PEDOT:PSS, the α value is 0.83 whereas for NiO it is 0.95. The α value close to 1 for NiO indicates that no space charge builds up and that photocurrent extraction is efficient. In contrast the value for PEDOT:PSS indicates that, the photocurrent extraction is substantially space charge limited. Thus in NiO HEL based perovskite solar cells, the photocurrent extraction is very efficient whereas in PEDOT:PSS HEL based devices photocurrent extraction is space charge limited. This can account for the higher short circuit current density observed in NiO HEL based perovskite solar cells in comparison to that of PEDOT:PSS HELs. In order to understand the different recombination mechanisms prevalent in the devices, based on NiO and PEDOT:PSS HELs, the dependence of $V_{oc}$ on light intensity was investigated.

The dependence of open circuit voltage $V_{oc}$ on incident light intensity $I$ is given by the relation:

$$V_{oc} = \frac{nk_BT}{q} \ln I$$

  

where $n$ is the recombination ideality factor and $\frac{k_B T}{q}$ the thermal voltage. When the slope of $V_{oc}$ vs the logarithm of light intensity $I$ is $\frac{k_B T}{q}$, (corresponding to $n=1$) it indicates bimolecular recombination, whereas a steeper slope ($n>1$) indicates trap assisted recombination. Figure 4(b) shows that by linearly fitting the $V_{oc}$ vs light intensity a slope of 1.14 $k_B T/q$ and 2.0 $k_B T/q$ is obtained for NiO and PEDOT:PSS HEL based perovskite solar cells respectively. This indicates that, in the case of PEDOT:PSS based perovskite solar cells, trap assisted recombination is dominant whereas in the NiO based devices, Shockley-Read-Hall (SRH) recombination is largely suppressed and the recombination mechanism is mainly bimolecular. Variation of FF as a function of light intensity for both NiO and PEDOT:PSS based $CH_3NH_3PbI_3$ perovskite solar cells is given in Figure S4. At low light intensity, in the case of NiO the FF is higher whereas in the case of PEDOT:PSS, the FF systematically decreases with lowering the light intensity. This behavior of FF with light intensity also supports the lower recombination losses in NiO compared to that in PEDOT:PSS HELs[35, 36].

### 3.3 Properties of $CH_3NH_3PbI_3$ films on PEDOT:PSS and NiO Hole extraction layers:

To understand the origin of this difference in recombination dynamics of photogenerated charge carriers in solar cells, based on NiO and PEDOT:PSS as HELs, the structural, morphological and photophysical properties of the $CH_3NH_3PbI_3$ thin films on these interlayers were studied. Figure 4(c) shows the UV-VIS absorption spectra of the $CH_3NH_3PbI_3$ films on NiO and PEDOT:PSS interlayers and both show comparable absorbance over the entire wavelength range of 400-800 nm. This implies that the difference in short circuit current density observed in Figure 2(b) and Table 1, does not arise from the absorption or charge generation differences in $CH_3NH_3PbI_3$ film grown on NiO or PEDOT:PSS. To understand the crystalline properties of the perovskite layer grown on these different interlayers, the X-ray diffraction (XRD) pattern of the $CH_3NH_3PbI_3$ films were measured and are shown in Figure 4(d). XRD spectra of both perovskite films exhibit comparable intensities of diffraction signals at 14.1° and 28.5°, corresponding to the (110) and (220) planes of the tetragonal phase. No trace of $PbI_2$ peak was seen in the XRD spectra of perovskite films grown on either of the interlayers. Thus, the observed differences in photovoltaic performance shown in Figure 2(b) are not related to differences in absorption or crystalline properties of perovskite active layer. To understand the contribution of any morphological differences, scanning electron microscopy (SEM) images were taken. **Figure 5(a) and (b)** show the corresponding SEM

   10

images. As seen from the images, the $CH_3NH_3PbI_3$ perovskite layer grown on NiO consists of dense and uniform perovskite crystalline domains of size ranging from 100- 250 nm. In contrast the $CH_3NH_3PbI_3$ perovskite layer grown on PEDOT:PSS has a different morphology with non-uniform distribution of perovskite domain sizes. In addition to the non-homogeneous distribution, many aggregate features are also seen on the grain boundaries. Since $PbI_2$ related features were not seen in the XRD spectra, these small aggregates are also ascribed to perovskite crystals. These non-homogeneous and aggregate features can impede charge transport leading to recombination losses and space charge build up. This property can further account for the lower short circuit current density and lower $V_{oc}$ in the case of PEDOT:PSS HEL based solar cells.

To further investigate the hole extraction efficiency to the NiO and PEDOT:PSS layers, steady state and time resolved PL spectroscopy studies were carried out for the $CH_3NH_3PbI_3$ layers grown on these hole extraction layers as well as on perovskite layer grown on glass substrates. **Figure 5c** shows the normalised time integrated PL spectra of the $CH_3NH_3PbI_3$ layers grown on NiO and PEDOT:PSS HELs recorded using identical excitation intensity. Both samples show identical shape of PL spectra with a peak at 768 nm which indicates that perovskite layer has the same optical bandgap on different interlayers. The time resolved PL decays are shown in **Figure 5(d).** The PL in $CH_3NH_3PbI_3$ perovskite layer comes from bimolecular recombination and its intensity is proportional to the product of electron and hole densities. The perovskite layer grown on glass shows very little decay on the time scale of 2 ns which indicates that charge recombination is slow. PL decays on NiO and PEDOT are faster which indicates hole extraction from the perovskite layer. Faster hole extraction is observed into NiO which agrees with previous observations of more efficient hole extraction into NiO [37, 38]. The PL decay on NiO is nearly exponential with about 300 ps lifetime and shows that hole extraction occurs with a well-defined homogeneous rate. In contrast, the layer on PEDOT:PSS shows significant non-exponential decay with a long component exceeding 2 ns. This indicates that hole extraction to PEDOT:PSS is very heterogeneous; a fraction of holes are extracted on the same time scale as to NiO but other holes are extracted very slowly. The homogeneous rate of hole transfer from $CH_3NH_3PbI_3$ to NiO can be explained by better energy level alignment of $CH_3NH_3PbI_3$ with NiO (shown in Figure S5) and uniform and homogeneous perovskite morphology on NiO HEL compared to PEDOT:PSS HEL. In the case of $CH_3NH_3PbI_3$ layer on PEDOT:PSS there is unfavourable energy level alignment and non-uniform, aggregated morphology, hence holes can diffuse back into bulk perovskite and can

 11

take many attempts to get extracted. Faster hole extraction to NiO will also enhance charge collection rate and prevent charge accumulation at the interface of perovskite with the HEL. This faster collection of holes and the possibility for reduced recombination can thus account for the improved short circuit current (hence higher EQE) and better $V_{oc}$ of $CH_3NH_3PbI_3$ solar cells on NiO HEL compared to those on PEDOT:PSS HEL.

From the morphological studies and photophysical characterization, the higher short circuit current density of $CH_3NH_3PbI_3$ solar cells on NiO is revealed due to homogeneous morphology, better energy level alignment and faster hole extraction from the perovskite active layer. Even though the higher Voc of $CH_3NH_3PbI_3$ solar cells on NiO can be explained by less recombination losses and favourable energy alignment, to confirm this further surface photovoltage measurement is performed and explained below.

### *3.4 Surface photovoltage measurements*

To understand the contribution of HELs specifically to the open circuit voltage, surface photovoltage was measured using a macroscopic Kelvin probe set up. This method allows the measurement of photovoltage of partial or incomplete solar cell device architectures and analysis of each contact (buried interface) on the role of charge extraction and has previously been successfully used to study $CH_3NH_3PbI_3$ layers grown on different interlayers[39]. Previously Lee *et al*[40] have shown that surface photovoltage, which is directly related to the open circuit voltage depends on the relative efficiency of charge extraction at the interfaces. **Figure 6** shows the surface photovoltage measured for partial device structure of ITO/NiO or PEDOT:PSS/$CH_3NH_3PbI_3$ for PEDOT:PSS and NiO HELs. The surface photovoltage measured for NiO based structure is 160 meV whereas the surface photovoltage measured for PEDOT:PSS is only 70 meV. Thus ~ 90 meV higher surface photovoltage for the ITO/NiO/$CH_3NH_3PbI_3$ heterostructure in comparison to ITO/ PEDOT:PSS/$CH_3NH_3PbI_3$ implies better hole extraction through NiO.

After realising efficient $CH_3NH_3PbI_3$ perovskite solar cells on NiO HEL and the superior quality of the $CH_3NH_3PbI_3$ perovskite layers grown on NiO, the fabrication of mixed halide perovskite solar cells on NiO was attempted as described in the next section.

### *3.5 NiO HEL extended to other mixed halide perovskite solar cells*

With the success of fabrication of efficient $CH_3NH_3PbI_3$ based solar cells in inverted architecture, mixed halides such as $CH_3NH_3PbI_{2.9}Cl_{0.1}$ and $CH_3NH_3PbI_{2.9}Br_{0.1}$ based perovskite solar cells were also fabricated using NiO as HEL. The $CH_3NH_3PbI_{2.9}Cl_{0.1}$ based solar cells show a maximum PCE of 12.5 % and for $CH_3NH_3PbI_{2.9}Br_{0.1}$ based solar cells the maximum PCE is 11.3 %. The corresponding J-V characteristics, EQE spectra and the photovoltaic performance parameters are shown in Figure S6 and Table S2. The high efficiency of mixed halide-based perovskite solar cells implies the wide applicability of the NiO HEL to a range of perovskite solar cells.

So far, the properties of NiO were compared to the most commonly employed HEL of PEDOT:PSS and the superior photovoltaic performance of NiO HEL based $CH_3NH_3PbI_3$ and mixed halide perovskite solar cells under 1 Sun illumination was demonstrated and explained. In the next section, the indoor photovoltaic performance of NiO based perovskite solar cells is investigated.

### 3.6 Indoor photovoltaic testing for hybrid perovskites:

Hybrid perovskite solar cells are primarily tested and certified for outdoor applications and record efficiencies are reported for illumination at 1 Sun. Compared to the outdoor photovoltaic performance of hybrid perovskites, little attention has been given to their indoor photovoltaic properties. It has been previously reported that in the case of indoor lighting conditions, the optimum material bandgap is 1.9-2.0 eV in order to maximise the power conversion efficiency[41]. Hence mixed halide perovskites, as used here are more suitable than $CH_3NH_3PbI_3$ because of their larger bandgap. According to a previous report by Steim *et al* [42] the role of interlayers is crucial in determining the indoor photovoltaic performance: the interlayers should have higher shunt resistance and low series resistance. The low temperature solution processed NiO interlayer-based hybrid perovskites, which showed promising performance at AM 1.5 G solar spectrum, have been studied for their indoor photovoltaic performance without further optimization. White LED and fluorescent lamps were used as the indoor light sources. **Figure 7 and Table 3** show the J-V characteristics and the photovoltaic performance of $CH_3NH_3PbI_3$, and mixed halide perovskites as a function of white LED light with intensity 2.2 mW/cm$^2$. Compared to pure $CH_3NH_3PbI_3$ solar cells, the mixed halide perovskite solar cells outperform under white LED lighting at very low intensities. Under low light intensity, the PCE is enhanced almost 100 % for $CH_3NH_3PbI_{2.9}Cl_{0.1}$ mixed halide perovskite solar cells, giving rise to 20.8 % and for $CH_3NH_3PbI_{2.9}Br_{0.1}$ the PCE is 19.9 %. This could be related to the slightly higher bandgap of mixed halide perovskites (1.63 vs 1.57 eV) compared to pure $CH_3NH_3PbI_3$ giving rise to the better spectral overlap of solar cell EQE spectra with the indoor LED emission spectra [Figure 7(b)]. As can be seen from Table 3, this higher bandgap also leads to higher $V_{oc}$ for mixed halide perovskite solar cells compared to $CH_3NH_3PbI_3$ solar cells. In addition, the improved charge carrier diffusion length reported in chloride based mixed halide compared to bromide and pure iodide-based perovskite can also contribute to this improved device performance [43]. In all the cases the high FF of the solar cells under indoor lighting imply the better hole extraction through the NiO layer.

After measuring the performance of NiO HEL based $CH_3NH_3PbI_3$ and mixed halide perovskites solar cells for white LED lighting, their performance was characterized under fluorescent lamp. The fluorescent lamp spectrum is shown in **Figure 7(b)**. The incident light intensity used was 0.32 mW/cm$^2$. The J-V characteristics and the corresponding photovoltaic performance parameters are shown in **Figure 8 and Table 4**. Similar to the performance under white LED, mixed halide perovskite solar cells of the $CH_3NH_3PbI_{2.9}Cl_{0.1}$ show the highest performance of 23.0 %. The high FF of these solar cells even under very low intensity confirms the superior charge selection properties of the NiO nanoparticle interlayers.

Up to this point, the performance of NiO nanoparticle thin films as efficient hole extraction layers in $CH_3NH_3PbI_3$, $CH_3NH_3PbI_{2.9}Br_{0.1}$, $CH_3NH_3PbI_{2.9}Cl_{0.1}$ based solar cells were demonstrated for AM 1.5 G solar spectrum and indoor lighting condition of white LED and fluorescent light. Now finally, the stability of the NiO nanoparticle HEL based solar cells and that of PEDOT:PSS HEL is investigated.

### 3.7 Shelf- life stability studies

For stability studies, the encapsulated solar cells were kept in the dark under ambient air conditions of temperature 22 °C and relative humidity 40% (measured using HTC-1 thermo-hygrometer). The shelf -life study (3 months and 20 days, encapsulated solar cells kept under above mentioned ambient conditions) of the $CH_3NH_3PbI_3$ based solar cells fabricated on NiO and PEDOT:PSS HELs shows that the degradation of the hybrid perovskite is faster on PEDOT:PSS HEL with much lower yield (12%- 1 out of 8 pixels) after the degradation, compared to the solar cells on NiO HEL (75 %- 6 out of 8 pixels). **Figure 9(a)** shows the appearance of NiO and PEDOT:PSS based $CH_3NH_3PbI_3$ solar cells after 3 months and 20 days. The yellowish appearance of the $CH_3NH_3PbI_3$ on PEDOT:PSS indicates moisture induced dissociation of the perovskite active layer whereas the $CH_3NH_3PbI_3$ films on NiO HEL retains its black brown perovskite appearance. This implies better moisture permeation prevention for the NiO based HEL compared to the PEDOT:PSS [44]. Regarding the photovoltaic performance of $CH_3NH_3PbI_3$ perovskite solar cells, both NiO and PEDOT:PSS HEL based devices retain 60-65% of initial PCE. The corresponding photovoltaic performance parameters are listed in the **Table S3**. However in the case of mixed halide perovskite solar cells $CH_3NH_3PbI_{2.9}Br_{0.1}$, $CH_3NH_3PbI_{2.9}Cl_{0.1}$, a better shelf-life stability of above 90 % is obtained. A summary of the stability study is given in **Figure 9(b)** and the detailed photovoltaic performance parameters are listed in Table S4 and in Figure S7.

4. Conclusion

We have demonstrated that simple solution-processed NiO thin films are efficient hole extraction layers for perovskite solar cells. In particular we find that NiO leads to much better performance than PEDOT:PSS in *p-i-n* perovskite solar cells. Under solar illumination we obtained a power conversion efficiency of 13.3% for the NiO containing solar cells compared with 7.9% for those made with PEDOT:PSS. We attribute this improvement to superior microstructural, optoelectronic and interface properties of hybrid perovskites grown on NiO. In addition we showed that the NiO layer worked well with mixed halide perovskites as well as for $CH_3NH_3PbI_3$.

The low temperature processing of NiO means that we could readily make devices on flexible substrates. We also explored the photovoltaic performance under indoor lighting. We found that the low recombination rate and fast extraction of photogenerated carriers by NiO is particularly suitable for low intensity indoor illumination and obtained a power conversion efficiency of 23%. Further attractive features of using a NiO hole extraction layer are that it acts as a barrier to moisture and so enhances shelf-life. These good results for a range of hybrid perovskites, substrates and lighting conditions indicate that NiO nanoparticle thin films are promising for the future development of perovskite solar cells and other optoelectronic devices.

ACKNOWLEDGEMENTS

We are grateful to the European Commission for financial support through the grant, EXCITON 321305. Dr. L.K.Jagadamma acknowledges support from a Marie Skłodowska-Curie Individual Fellowship (European Commission) (MCIF: No. 745776). We are grateful to EPSRC for equipment grant (EP/L017008/1). The authors are grateful to Dr. J. Payne for performing X-ray diffraction measurements and Dr. D. N. Miller for the transmission electron microscopy measurements.

**Figure Captions**

**Figure 1:** (a) Comparison of the transmission spectra of NiO nanoparticle thin films with ITO substrate and PEDOT:PSS. (b) AFM image of NiO nanoparticle thin film (c) TEM image of NiO nanoparticle thin film. Inset shows the selective area diffraction pattern (d) Ambient photoemission spectra of NiO nanoparticle thin films on ITO substrate.

**Figure 2:** (a) Inverted perovskite solar cell device architecture (b) J-V characteristics comparison of the best perovskite solar cells on NiO and PEDOT: PSS HELs. RW stands for reverse bias scanning and FW stands for forward bias scanning. (c) Histogram of solar cell device numbers and power conversion efficiency distribution for NiO and PEDOT: PSS HELs (d) EQE spectra comparison of the best solar cell devices for NiO and PEDOT: PSS HELs .

**Figure 3:** (a) J-V characteristics of the $CH_3NH_3PbI_3$ perovskite solar cells on flexible PET substrate using NiO HEL. RW stands for reverse bias scanning and FW stands for forward bias scanning. (b) The corresponding EQE spectra of the $CH_3NH_3PbI_3$ perovskite solar cell. Inset shows the fabricated flexible perovskite solar cells.

**Figure 4:** Light intensity dependence of perovskite solar cells based on NiO and PEDOT:PSS HELs for (a) $J_{sc}$ (log-log scale) and (b) $V_{oc}$ (semi log scale). (c) Absorbance spectra and (d) X-ray diffraction pattern comparison for $CH_3NH_3PbI_3$ layers grown on PEDOT:PSS and NiO HELs

**Figure 5:** Scanning electron microscopy images of $CH_3NH_3PbI_3$ perovskite layer on (a) NiO and (b) PEDOT:PSS HELs. (c) Normalised time integrated PL spectra (over 2 ns) and (d) PL decays of $CH_3NH_3PbI_3$ layers grown on glass without hole extraction layers, and with NiO and PEDOT:PSS HELs. Solid lines represent the bi-exponential fit.

**Figure 6:** (a) Schematic of the incomplete device architecture used for measuring the surface photovoltage. (b) surface photovoltage measured for NiO as HEL and (c) surface photovoltage measured for PEDOT:PSS as HEL.

**Figure 7:** (a) J-V characteristics of the NiO HEL based $CH_3NH_3PbI_3$ and mixed halide perovskite solar cells under white light LED (2.20 mW/cm$^2$). (b) White light LED and fluorescent lamp spectral overlap with the EQE of different perovskite solar cells.

**Figure 8:** (a) J-V characteristics of the NiO HEL based $CH_3NH_3PbI_3$ and mixed halide perovskite solar cells under fluorescent lamp (0.32 mW/cm$^2$).

**Figure 9:** (a) Images showing the NiO and PEDOT:PSS HELs based $CH_3NH_3PbI_3$ perovskite solar cells after 3 months and 20 days of shelf life under ambient conditions. (b) Summary of the shelf-life studies of the various perovskite solar cells on NiO HEL.

## References

[1] NREL chart, in, June 2019.
[2] T. Salim, S. Sun, Y. Abe, A. Krishna, A.C. Grimsdale, Y.M. Lam, Perovskite-based solar cells: impact of morphology and device architecture on device performance, Journal of Materials Chemistry A, 3 (2015) 8943-8969.
[3] P. Gao, M. Grätzel, M.K. Nazeeruddin, Organohalide lead perovskites for photovoltaic applications, Energy & Environmental Science, 7 (2014) 2448-2463.

[4] K. Domanski, J.-P. Correa-Baena, N. Mine, M.K. Nazeeruddin, A. Abate, M. Saliba, W. Tress, A. Hagfeldt, M. Grätzel, Not All That Glitters Is Gold: Metal-Migration-Induced Degradation in Perovskite Solar Cells, ACS Nano, 10 (2016) 6306-6314.
[5] L. Meng, J. You, T.-F. Guo, Y. Yang, Recent Advances in the Inverted Planar Structure of Perovskite Solar Cells, Accounts of Chemical Research, 49 (2016) 155-165.
[6] T.A. Berhe, W.-N. Su, C.-H. Chen, C.-J. Pan, J.-H. Cheng, H.-M. Chen, M.-C. Tsai, L.-Y. Chen, A.A. Dubale, B.-J. Hwang, Organometal halide perovskite solar cells: degradation and stability, Energy & Environmental Science, 9 (2016) 323-356.
[7] X.C. Yang, H.X. Wang, B. Cai, Z. Yu, L.C. Sun, Progress in hole-transporting materials for perovskite solar cells, Journal of Energy Chemistry, 27 (2018) 650-672.
[8] M.H. Li, J.H. Yum, S.J. Moon, P. Chen, Inorganic p-Type Semiconductors: Their Applications and Progress in Dye-Sensitized Solar Cells and Perovskite Solar Cells, Energies, 9 (2016) 28.
[9] M.B. Islam, M. Yanagida, Y. Shirai, Y. Nabetani, K. Miyano, NiOx Hole Transport Layer for Perovskite Solar Cells with Improved Stability and Reproducibility, ACS Omega, 2 (2017) 2291-2299.
[10] B. Groeneveld, M. Najafi, B. Steensma, S. Adjokatse, H.H. Fang, F. Jahani, L. Qiu, G.H. ten Brink, J.C. Hummelen, M.A. Loi, Improved efficiency of NiOx-based p-i-n perovskite solar cells by using PTEG-1 as electron transport layer, Apl Materials, 5 (2017).
[11] Z.J. Hu, D. Chen, P. Yang, L.J. Yang, L.S. Qin, Y.X. Huang, X.C. Zhao, Sol-gel-processed yttrium-doped NiO as hole transport layer in inverted perovskite solar cells for enhanced performance, Applied Surface Science, 441 (2018) 258-264.
[12] Y. Wei, K. Yao, X.F. Wang, Y.H. Jiang, X.Y. Liu, N.G. Zhou, F. Li, Improving the efficiency and environmental stability of inverted planar perovskite solar cells via silver-doped nickel oxide hole-transporting layer, Applied Surface Science, 427 (2018) 782-790.
[13] J. Tang, D. Jiao, L. Zhang, X.Z. Zhang, X.X. Xu, C. Yao, J.H. Wu, Z. Lan, High-performance inverted planar perovskite solar cells based on efficient hole-transporting layers from well-crystalline NiO nanocrystals, Solar Energy, 161 (2018) 100-108.
[14] Z.L. Zhu, Y. Bai, T. Zhang, Z.K. Liu, X. Long, Z.H. Wei, Z.L. Wang, L.X. Zhang, J.N. Wang, F. Yan, S.H. Yang, High-Performance Hole-Extraction Layer of Sol-Gel-Processed NiO Nanocrystals for Inverted Planar Perovskite Solar Cells, Angewandte Chemie-International Edition, 53 (2014) 12571-12575.
[15] M.-H. Liu, Z.-J. Zhou, P.-P. Zhang, Q.-W. Tian, W.-H. Zhou, D.-X. Kou, S.-X. Wu, p-type Li, Cu-codoped NiOx hole-transporting layer for efficient planar perovskite solar cells, Opt. Express, 24 (2016) A1349-A1359.
[16] K.-C. Wang, J.-Y. Jeng, P.-S. Shen, Y.-C. Chang, E.W.-G. Diau, C.-H. Tsai, T.-Y. Chao, H.-C. Hsu, P.-Y. Lin, P. Chen, T.-F. Guo, T.-C. Wen, p-type Mesoscopic Nickel Oxide/Organometallic Perovskite Heterojunction Solar Cells, Scientific Reports, 4 (2014) 4756.
[17] J.-Y. Jeng, K.-C. Chen, T.-Y. Chiang, P.-Y. Lin, T.-D. Tsai, Y.-C. Chang, T.-F. Guo, P. Chen, T.-C. Wen, Y.-J. Hsu, Nickel Oxide Electrode Interlayer in CH3NH3PbI3 Perovskite/PCBM Planar-Heterojunction Hybrid Solar Cells, Advanced Materials, 26 (2014) 4107-4113.
[18] W. Zhang, X. Zhang, T. Wu, W. Sun, J. Wu, Z. Lan, Interface engineering with NiO nanocrystals for highly efficient and stable planar perovskite solar cells, Electrochimica Acta, 293 (2019) 211-219.
[19] W. Zhang, J. Song, D. Wang, K. Deng, J. Wu, L. Zhang, Dual interfacial modification engineering with p-type NiO nanocrystals for preparing efficient planar perovskite solar cells, Journal of Materials Chemistry C, 6 (2018) 13034-13042.
[20] K. Deevi, V.S.R. Immareddy, Synthesis and characterization of optically transparent nickel oxide nanoparticles as a hole transport material for hybrid perovskite solar cells, Journal of Materials Science: Materials in Electronics, 30 (2019) 6242-6248.
[21] Z. Liu, A. Zhu, F. Cai, L. Tao, Y. Zhou, Z. Zhao, Q. Chen, Y.-B. Cheng, H. Zhou, Nickel oxide nanoparticles for efficient hole transport in p-i-n and n-i-p perovskite solar cells, Journal of Materials Chemistry A, 5 (2017) 6597-6605.

[22] W.A. MacDonald, M.K. Looney, D. MacKerron, R. Eveson, R. Adam, K. Hashimoto, K. Rakos, Latest advances in substrates for flexible electronics, Journal of the Society for Information Display, 15 (2007) 1075-1083.
[23] Q. He, K. Yao, X. Wang, X. Xia, S. Leng, F. Li, Room-Temperature and Solution-Processable Cu-Doped Nickel Oxide Nanoparticles for Efficient Hole-Transport Layers of Flexible Large-Area Perovskite Solar Cells, ACS Applied Materials & Interfaces, 9 (2017) 41887-41897.
[24] W. Chen, Y. Wu, J. Fan, A.B. Djurišić, F. Liu, H.W. Tam, A. Ng, C. Surya, W.K. Chan, D. Wang, Z.-B. He, Understanding the Doping Effect on NiO: Toward High-Performance Inverted Perovskite Solar Cells, Adv. Energy Mater., 8 (2018) 1703519.
[25] C.-Y. Chen, J.-H. Chang, K.-M. Chiang, H.-L. Lin, S.-Y. Hsiao, H.-W. Lin, Perovskite Photovoltaics for Dim-Light Applications, Advanced Functional Materials, 25 (2015) 7064-7070.
[26] J. Dagar, S. Castro-Hermosa, G. Lucarelli, F. Cacialli, T.M. Brown, Highly efficient perovskite solar cells for light harvesting under indoor illumination via solution processed SnO2/MgO composite electron transport layers, Nano Energy, 49 (2018) 290-299.
[27] G. Lucarelli, F. Di Giacomo, V. Zardetto, M. Creatore, T.M. Brown, Efficient light harvesting from flexible perovskite solar cells under indoor white light-emitting diode illumination, Nano Research, 10 (2017) 2130-2145.
[28] H.K.H. Lee, J. Barbé, S.M.P. Meroni, T. Du, C.-T. Lin, A. Pockett, J. Troughton, S.M. Jain, F. De Rossi, J. Baker, M.J. Carnie, M.A. McLachlan, T.M. Watson, J.R. Durrant, W.C. Tsoi, Outstanding Indoor Performance of Perovskite Photovoltaic Cells – Effect of Device Architectures and Interlayers, Solar RRL, 3 (2019) 1800207.
[29] M. Li, C. Zhao, Z.-K. Wang, C.-C. Zhang, H.K.H. Lee, A. Pockett, J. Barbé, W.C. Tsoi, Y.-G. Yang, M.J. Carnie, X.-Y. Gao, W.-X. Yang, J.R. Durrant, L.-S. Liao, S.M. Jain, Interface Modification by Ionic Liquid: A Promising Candidate for Indoor Light Harvesting and Stability Improvement of Planar Perovskite Solar Cells, Advanced Energy Materials, 8 (2018) 1801509.
[30] E.L. Unger, L. Kegelmann, K. Suchan, D. Sorell, L. Korte, S. Albrecht, Roadmap and roadblocks for the band gap tunability of metal halide perovskites, Journal of Materials Chemistry A, 5 (2017) 11401-11409.
[31] M. Kumano, M. Ide, N. Seiki, Y. Shoji, T. Fukushima, A. Saeki, A ternary blend of a polymer, fullerene, and insulating self-assembling triptycene molecules for organic photovolatics, Journal of Materials Chemistry A, 4 (2016) 18490-18498.
[32] J.A. Azurdia, J. Marchal, P. Shea, H. Sun, X.Q. Pan, R.M. Laine, Liquid-Feed Flame Spray Pyrolysis as a Method of Producing Mixed-Metal Oxide Nanopowders of Potential Interest as Catalytic Materials. Nanopowders along the NiO–Al2O3 Tie Line Including (NiO)0.22(Al2O3)0.78, a New Inverse Spinel Composition, Chemistry of Materials, 18 (2006) 731-739.
[33] L.J.A. Koster, V.D. Mihailetchi, R. Ramaker, P.W.M. Blom, Light intensity dependence of open-circuit voltage of polymer:fullerene solar cells, Applied Physics Letters, 86 (2005) 123509.
[34] S.R. Cowan, A. Roy, A.J. Heeger, Recombination in polymer-fullerene bulk heterojunction solar cells, Physical Review B, 82 (2010) 245207.
[35] J. You, L. Dou, K. Yoshimura, T. Kato, K. Ohya, T. Moriarty, K. Emery, C.-C. Chen, J. Gao, G. Li, Y. Yang, A polymer tandem solar cell with 10.6% power conversion efficiency, Nature Communications, 4 (2013) 1446.
[36] E. Karimi, S.M.B. Ghorashi, Simulation of perovskite solar cell with P$_3$HT hole-transporting materials, in, SPIE, 2017, pp. 14.
[37] E. Bi, H. Chen, F. Xie, Y. Wu, W. Chen, Y. Su, A. Islam, M. Grätzel, X. Yang, L. Han, Diffusion engineering of ions and charge carriers for stable efficient perovskite solar cells, Nature Communications, 8 (2017) 15330.
[38] L.-B. Huang, P.-Y. Su, J.-M. Liu, J.-F. Huang, Y.-F. Chen, S. Qin, J. Guo, Y.-W. Xu, C.-Y. Su, Interface engineering of perovskite solar cells with multifunctional polymer interlayer toward improved performance and stability, Journal of Power Sources, 378 (2018) 483-490.

[39] J.R. Harwell, T.K. Baikie, I.D. Baikie, J.L. Payne, C. Ni, J.T.S. Irvine, G.A. Turnbull, I.D.W. Samuel, Probing the energy levels of perovskite solar cells via Kelvin probe and UV ambient pressure photoemission spectroscopy, Physical Chemistry Chemical Physics, 18 (2016) 19738-19745.

[40] L. Barnea-Nehoshtan, S. Kirmayer, E. Edri, G. Hodes, D. Cahen, Surface Photovoltage Spectroscopy Study of Organo-Lead Perovskite Solar Cells, The Journal of Physical Chemistry Letters, 5 (2014) 2408-2413.

[41] M. Freunek, M. Freunek, L.M. Reindl, Maximum efficiencies of indoor photovoltaic devices, IEEE Journal of Photovoltaics, 3 (2013) 59-64.

[42] R. Steim, T. Ameri, P. Schilinsky, C. Waldauf, G. Dennler, M. Scharber, C.J. Brabec, Organic photovoltaics for low light applications, Solar Energy Materials and Solar Cells, 95 (2011) 3256-3261.

[43] S.D. Stranks, G.E. Eperon, G. Grancini, C. Menelaou, M.J.P. Alcocer, T. Leijtens, L.M. Herz, A. Petrozza, H.J. Snaith, Electron-Hole Diffusion Lengths Exceeding 1 Micrometer in an Organometal Trihalide Perovskite Absorber, Science, 342 (2013) 341-344.

[44] Y. Hu, J. Schlipf, M. Wussler, M.L. Petrus, W. Jaegermann, T. Bein, P. Müller-Buschbaum, P. Docampo, Hybrid Perovskite/Perovskite Heterojunction Solar Cells, ACS Nano, 10 (2016) 5999-6007.